\documentstyle[aps,prl,multicol,rotate,psfig]{revtex}
\begin{document}
\draft

\title{Effect of Gravity and Confinement on Phase Equilibria: \\
A Density Matrix Renormalization Approach}
\author{Enrico Carlon$^{1}$ and Andrzej Drzewi\'nski$^{1,2}$}
\address{$^1$ Institute for Theoretical Physics, Katholieke Universiteit
Leuven, Celestijnenlaan 200D, B-3001 Leuven, Belgium}
\address{$^2$ Institute for Low Temperature and Structure Research, 
Polish Academy of Sciences,\\
 P.O.Box 937, 50-950 Wroc\l aw 2, Poland}
\maketitle

\begin{abstract}
The effect of gravity on a two dimensional fluid, or Ising magnet, confined 
between opposing walls is analyzed by density matrix renormalization. Gravity 
restores two phase coexistence up to the bulk critical point, in agreement 
with mean field calculations. A detailed finite size scaling analysis of the
critical point shift is performed. Density matrix renormalization results are 
very accurate and the technique is promising and best suited to study equilibrium 
properties of two dimensional classical systems in contact with walls or with 
free surfaces.
\end{abstract}


\begin{multicols}{2} \narrowtext
The thermodynamical properties of confined systems have received 
a lot of attention in the past years 
\cite{fisher,parry,parry2,jos,albano,albano2,binder,binder2,fisher2,maciolek}. 
Their critical behavior is rather different from the bulk criticality 
and has been the subject of
extended investigations by means of mean field and scaling analysis
\cite{fisher,parry,parry2,jos}, 
Monte Carlo simulations \cite{albano,albano2,binder,binder2} 
and exact calculations \cite{fisher2,maciolek}.
The simplest and most studied case is the Ising model in a $L \times 
M^{d-1}$ lattice with $M \to \infty$, i.e. confined between two 
infinite walls separated by a finite distance $L$. 
Of considerable interest is the situation in which magnetic fields
($h_1$ and $h_2$) acting on the spins at the walls are introduced. 

\begin{figure}[b]
\centerline{
\rotate[r]{\psfig{file=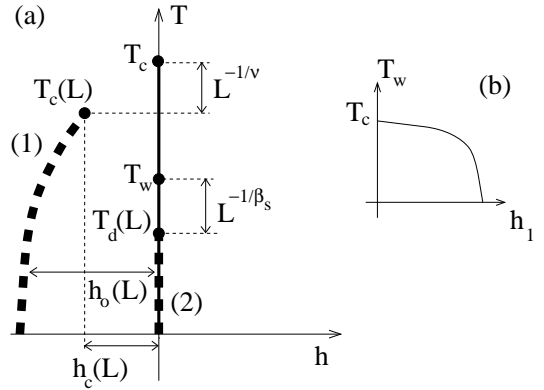,height=7cm}}}
\vskip 0.2truecm
\caption{(a) Phase diagram of the d-dimensional Ising model for a bulk 
system in the (h,T) plane (solid line). The dashed lines are the phase 
diagrams of confined systems with identical (1) and opposing (2) 
surface fields. (b) Dependence of the wetting temperature on the 
surface field $h_1$.}
\label{FIG01}
\end{figure}

For {\em parallel} surface fields ($h_1 \cdot h_2 > 0$) and finite 
$L$ two phase coexistence is shifted to finite values of
a bulk magnetic field $h$ \cite{fisher}, as illustrated in 
Fig.\ \ref{FIG01}(a). 
This phenomenon is analogous to the capillary condensation for a fluid 
confined between two parallel surfaces, where the gas-liquid transition 
occurs at a lower pressure than in the bulk.
Finite size scaling \cite{fisher} predicts that the capillary critical 
point $[h_c(L),T_c(L)]$ scales as:
\begin{eqnarray}
T_c(L) - T_c \sim L^{-y_T} \,\,\,\,\,\,\,  
&\mbox{and}& \,\,\,\,\,\,\,  
h_c(L) \sim L^{-y_H},
\label{capill}
\end{eqnarray}
where $T_c$ is the bulk critical temperature and $y_T=1/\nu$,  
$y_H = d - \beta/\nu$ are the thermal and magnetic exponents, respectively
(here $d$ is the dimensionality, $\nu$ and $\beta$ are the correlation 
length and magnetization exponents, respectively). 

At fixed $T < T_c$ and finite $L$ the scaling to the first order 
line is of type \cite{fisher}:
\begin{eqnarray}
h_0(L) \sim 1/L.
\label{kelvin}
\end{eqnarray}
While the previous relation  has been verified in Monte Carlo simulations in 
$d=2,3$ \cite{albano2,binder}, a direct verification of the scaling of the 
capillary critical point has not been attempted yet, due to the high 
computational effort \cite{binder} needed to locate accurately $[h_c(L),T_c(L)]$.

The case of {\em opposing} surface fields (for simplicity we consider $h_1 = 
- h_2$) was analyzed in detail by Parry and Evans \cite{parry} using a 
Ginzburg-Landau approach. They found that two phase coexistence is restricted 
to temperatures below the {\em interface delocalization temperature} $T_d(L)$ 
as shown in Fig.\ \ref{FIG01}(a); the surprising result \cite{parry} is that 
$T_d(L)$ does not 
scale to the bulk point $T_c$, for $L \to \infty$, but to the wetting 
temperature $T_w$ as:
\begin{eqnarray}
T_d(L) - T_w \sim L^{-1/\beta_s}.
\label{wet}
\end{eqnarray}
Here $\beta_s$ is the exponent describing the divergence of the thickness 
of the wetting layer for a semi infinite system: $l \sim (T_w - T)^{-\beta_s}$.
We recall also that $T_w$ depends on the value of the surface field $h_1$ and
can be far away from the bulk critical point (see Fig.\ \ref{FIG01}(b)).
For $T_d(L) \leq T < T_c$ there is a single phase \cite{parry}, with 
an interface meandering freely between the walls.
Numerical results from extensive Monte Carlo simulations \cite{albano,binder2} 
in $d=2,3$ confirm the mean field scenario.

Rogiers and Indekeu \cite{jos} extended the model with opposing surface fields, 
including a bulk field which varies linearly along the finite direction of the 
system and models the effect of gravity on a confined fluid. They considered
the following Hamiltonian:
\end{multicols} \widetext
\begin{eqnarray}
H &=& -J \sum_{(i,j)} s_{(i,j)} s_{(i+1,j)} -J \sum_{(i,j)} s_{(i,j)} 
s_{(i,j+1)} + h_1 \sum_{j} s_{(1,j)} - h_1 \sum_{j} s_{(L,j)} + 
g \sum_{i=1}^L (2 i-1- L) \sum_{j} s_{(i,j)},
\label{ham}
\end{eqnarray}
\begin{multicols}{2} \narrowtext
where $J > 0$, $s_{(i,j)}=\pm 1$ and $1 \leq i \leq L$ labels the lattice sites 
along the finite direction, while $j$ labels the remaining 
$d-1$ directions. 
The parameter $g$ is the equivalent of the gravitational constant.
Using a mean field approach, it was found \cite{jos} that the ordinary 
scaling to the bulk critical point is restored in an extended parameter space 
where $g$ is included. 
While the mean field analysis is correct at dimensions higher than the upper 
critical dimension ($d=4$), its validity in lower dimensions, where thermal 
fluctuations become important, is questionable.

The aim of this Letter is to study the phase diagram of the model described
by the Hamiltonian (\ref{ham}) for an $L \times \infty$ strip, beyond the
mean field approximation. We test the validity of the conclusions of 
Ref. \cite{jos} at the lower critical dimension ($d=2$) and clarify the
mechanism of the critical point shift for a system confined between 
opposing walls. 
We analyze also the exponents governing this shift and test the validity
of scaling assumptions that have been proposed some time ago 
\cite{fisher,vanleeuwen}, but, to our knowledge, never derived directly 
from a microscopic model.

The numerical calculations are based on a density matrix renormalization 
group (DMRG) approach, a technique developed by White 
\cite{whitePRL,whitePRB} for the study of ground state properties of 
quantum spin chains. Nishino \cite{nishino} extended the DMRG to $d=2$ 
classical systems as a transfer matrix renormalization. 
Transfer matrices are frequently used for numerical investigations of 
$L \times \infty$ strips; computations are restricted to small strip 
widths since the dimension of the transfer matrix grows exponentially
with $L$ \cite{nota1}. In the DMRG algorithm one constructs effective 
transfer matrices of small dimensions (i.e. numerically tractable), but 
which describe strips of large widths.
The spin space is truncated in a very efficient manner and the numerical 
accuracy is very good \cite{whitePRL,whitePRB,nishino}, even for 
renormalized matrices of not too large dimensions \cite{nota2}.
More details will be presented elsewhere \cite{future}.

We start the analysis of the phase diagram of the model from its ground
state properties. At $T=0$, $h_1 < J$ and for a strip of width $L$
the ground state is double degenerate (two phase coexistence) with all 
spins either up or down for:
\begin{eqnarray}
\frac{4}{L^2} (h_1 -J ) \,\, \leq \,\, g \,\, \leq \,\, \frac{4}{L^2} (h_1 + J)
\label{ground}
\end{eqnarray}
For values of $g$ outside this interval the ground state is non degenerate,
with an infinitely long straight interface at the center of the strip  
separating a region with all spins up from a region with all spins down.

To distinguish the two phase coexistence region from a single 
interface-like state it is convenient to calculate the correlation function
between two neighboring spins at the center of the strip $c_{L/2}
\equiv \langle s_{\left(\frac{L}{2},j \right)} s_{\left( \frac{L}{2}+1,j
\right)}\rangle$.
At $T=0$, $c_{L/2}$ drops from $+1$ in the two phase coexistence 
region to $-1$ in the single phase region.
For nonzero temperatures, a sharp distinction between these two regions 
is not possible since true criticality does not occur for finite values of
$L$. However a peak in the temperature derivative of $c_{L/2}$ can be seen
as a pseudocritical point where the system goes through a smooth change
from two phase coexistence to a one phase state, without a true phase 
transition.
As the strip width increases these finite peaks shift towards the true
thermodynamic singularities of the bulk system. The analysis
of shifts of pseudocritical points has been considered already 
\cite{parry2,albano} for the present model in absence of gravity.

\begin{figure}[b]
\centerline{
\rotate[r]{\psfig{file=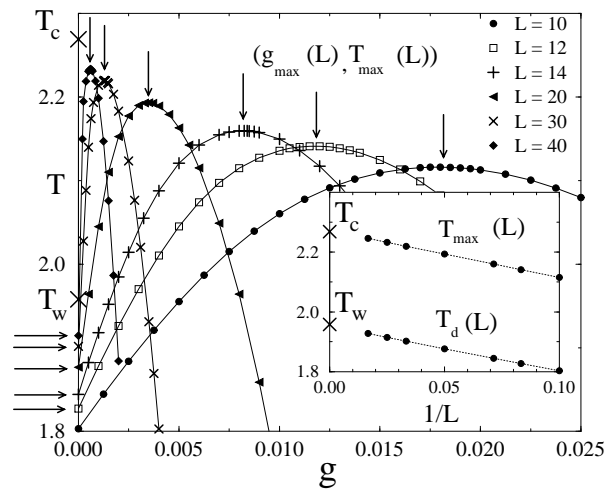,height=9cm}}}
\vskip 0.2truecm
\caption{Phase diagram of the model for different strip widths in the ($g,T$) plane
for $h_1=0.5$ and $J=1$; the area below the curves is the two phase coexistence region. 
Inset: Scaling of $T_{\rm{max}}(L)$ and $T_d(L)$ vs $1/L$. Error bars are much smaller
than symbol sizes.}
\label{FIG02}
\end{figure}

Figure\ \ref{FIG02} shows the phase diagram in the $(g,T)$ plane for $J=1$, 
$h_1=0.5$.
Each curve is the phase boundary between the two phase coexistence (area
below the curve) and the one phase region, for a specific value of $L$.
Only the phase boundaries for $L$ up to $40$ are shown although the largest
size considered is $L=60$.
For $g>0$ two phase coexistence is shifted to higher temperatures with respect 
to $g=0$ due to a competing effect of surface and bulk fields, while at negative
$g$ phase coexistence is suppressed. As $L$ increases the whole two phase 
coexistence region shrinks and shifts towards the $g=0$ axis. The intersections 
of the phase boundaries with this axis, indicated by the horizontal arrows in 
Fig.\ \ref{FIG02}, define the interface delocalization transition temperatures 
$T_d(L)$; the phase boundary maxima $[g_{\rm{max}}(L)$,$T_{\rm{max}}(L)]$ are 
indicated by the vertical arrows in the figure.

\begin{figure}[b]
\centerline{
\rotate[r]{\psfig{file=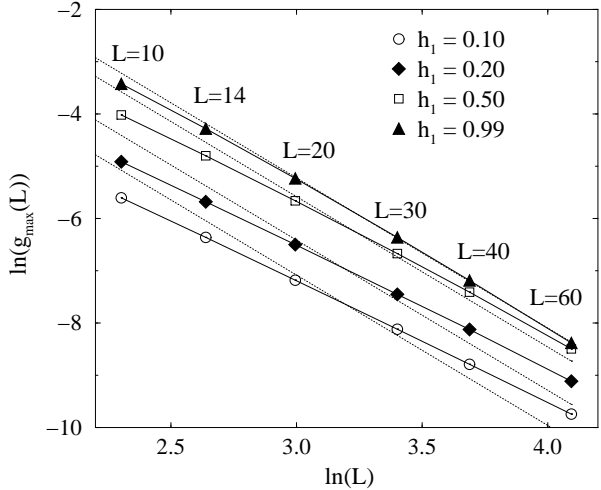,height=9cm}}}
\vskip 0.2truecm
\caption{Plot of $\ln[g_{\rm{max}}(L)]$ vs $\ln L$ for different surface fields. 
The dotted lines correspond to a slope $-2.875$. Error bars are smaller than
symbol sizes.}
\label{FIG03}
\end{figure}

We find for $T_d(L)$ a scaling in excellent agreement with Eq. (\ref{wet}) 
\cite{albano,binder2} and for $T_{\rm max}(L)$ a scaling of the type:
\begin{eqnarray}
T_{\rm max}(L) - T_c \sim L^{-y_T}
\label{newscal}
\end{eqnarray}
with the $d=2$ Ising exponents $\beta_s=1$ and $y_T=1$.
A plot of $T_{\rm{max}}(L)$ and $T_d(L)$ for $h_1=0.5$ as function of 
$1/L$ is shown in the inset of Fig.\ \ref{FIG02}. From an extrapolation of the data 
in the figure we find the following estimates for the wetting and bulk 
critical temperatures: $T_w = 1.9584(8)$ and $T_c = 2.272(4)$ (the exact
values are $T_w=1.95845$ \cite{abraham} and $T_c=2.269$). 
The same scaling analysis has been extended to several values of $h_1$: 
all data points are in excellent agreement with the relations (\ref{wet}) 
and (\ref{newscal}).

Van Leeuwen and Sengers \cite{vanleeuwen} analyzed the influence of gravity 
in an infinitely extended system; they argued that the product of $g$ with 
a length in the direction of the gravitational field should scale as a bulk 
constant field. On a finite strip, this leads to the following scaling
prediction \cite{jos}:
\begin{eqnarray}
g_{\rm{max}}(L) &\sim& L^{-\left( 1+y_H \right)}.
\label{gravscal}
\end{eqnarray}
Both at $T=0$ (see inequality (\ref{ground})) and also for finite
$T$ \cite{future}, not too close to the phase boundary maxima, we find a scaling
towards the bulk first order line of type $1/L^2$, in agreement with Van Leeuwen 
and Sengers' conjecture since the corresponding scaling of a constant bulk 
field is $1/L$ (see (\ref{kelvin})).

Figure\ \ref{FIG03} shows a plot of $\ln[g_{\rm{max}}(L)]$ vs $\ln(L)$ for four 
values of $h_1$, from $0.1$ to $0.99$ and $J=1$ \cite{nota3}. The dotted lines, 
drawn as a guide to the eye, correspond to a scaling of type (\ref{gravscal}) 
with the $d=2$ Ising exponent $1+y_H=2.875$. 
The data agree with the scaling relation (\ref{gravscal}) for the largest
surface field considered ($h_1 = 0.99$) for which a linear fit of the points 
for $L \geq 20$ yields an exponent $2.86(2)$ in good agreement with the 
Ising value. At lower surface fields and up to the largest size considered 
($L=60$) the exponent deviates from the expected value and increases from 
$2.4$ to $2.6$ as the surface field increases from $h_1=0.1$ to $h_1=0.5$.
The scaling analysis for $g_{\rm max}(L)$ is somewhat more problematic
than the scaling of $T_{\rm max}(L)$; this was to be expected since along 
the gravitational field direction there is a scaling to $T_c$ and a scaling
to the bulk first order line (of type $1/L^2$). The interplay between these
two scalings may be the cause of the observed shift of the exponent of
$g_{\rm max}(L)$ from the Ising value for low surface fields, where the
asymptotic behavior (\ref{gravscal}) possibly sets in for $L \gg 60$. 
In other studies of confined systems it was found that, in the scaling 
analysis of finite size data, the value of the system width beyond which 
one has a clear asymptotic behavior, may depend strongly on the surface 
field (see for instance Ref.\cite{binder}).

\begin{figure}[b]
\centerline{
\rotate[r]{\psfig{file=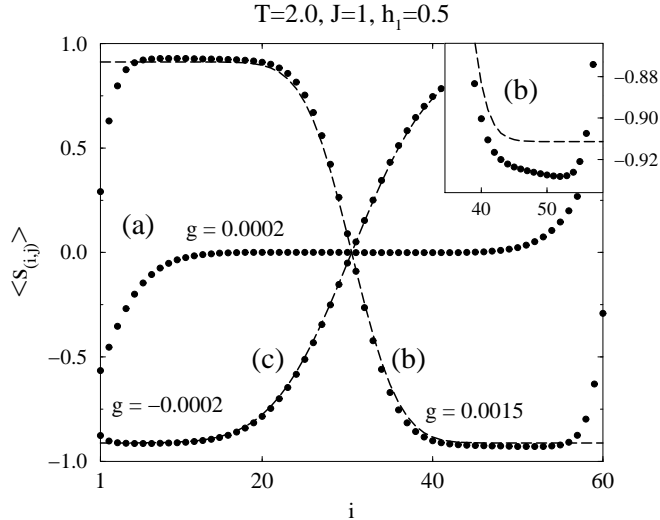,height=9cm}}}
\vskip 0.2truecm
\caption{(Circles) Magnetization profiles in the two phase coexistence 
region (a) and in the single phase region ((b) and (c)) calculated by
DMRG. (Dashed lines) Profiles calculated with a SOS approximation
[Eq. \protect{\ref{prof}}]. Inset: Detail of the profile (b) far from
the interfacial region.}
\label{FIG04}
\end{figure}

Figure\ \ref{FIG04} shows some magnetization profiles for $L=60$, $J=1$, 
$h_1=0.5$, $T=2.0$ and for different values of $g$. The profile (a) 
corresponds to a point of the phase diagram located in the two phase 
coexistence region, where the magnetization is averaged over two phases. 
This profile is similar to those calculated exactly by Macio\l ek and 
Stecki \cite{maciolek,nota4} for $g=0$ and $T < T_d(L)$. Profiles (b) 
and (c) correspond to points of the phase diagram in the one phase 
region. Notice the competing effect between gravity and surface fields 
in the vicinity of the walls in the profile (b).

Magnetization profiles of interface-like configurations can be
calculated using a solid-on-solid (SOS) approximation \cite{lipowsky}.
It is assumed that all the spins at the two sides of the interface are
fixed and take values $\pm m_0$; interfacial configurations with overhangs 
are neglected.

SOS magnetization profiles are shown in Fig.\ \ref{FIG04} (dashed lines) 
and are given by \cite{future}:
\begin{eqnarray}
\langle s_{(i,j)}\rangle = \frac{2 m_0}{\sqrt{\pi}} \,\, 
\int_0^{x(i)/\xi_\perp} dt \,\, e^{-t^2},
\label{prof}
\end{eqnarray}
where $x(i)=L[i-(L+1)/2]/(L-1)$ and $\xi_\perp$, the interface width, is given 
by the relation:
\begin{eqnarray}
\xi_\perp= \sqrt{\frac{T}{2 \sqrt{|g| m_0 \sigma_0}}}.
\end{eqnarray}
with $\sigma_0$ the surface tension. For $\sigma_0$ and $m_0$
we take their exactly known values at $g=0$; this approximation is good
for low gravity.

Results from the interface Hamiltonian agree well with those calculated with 
DMRG even at temperatures not too far from $T_c$ (in Fig.\ 4 the temperature 
is approximately $10\%$ below $T_c$). Gravity has the effect of reducing 
interface fluctuations, so that configurations with overhangs, which are not 
taken into account in the SOS model, have small weights.
We stress that (\ref{prof}) is valid in the limit $\xi_\perp \ll L$ since
the effect of the walls has been neglected.

In conclusion, we found that the competing effects of surface fields and 
gravity restore the ``ordinary" finite size scaling to the bulk critical 
point, in agreement with mean field predictions \cite{jos}, which thus survive 
the strong thermal fluctuations at the lower critical dimension.
It is currently believed that the model confined between opposing walls
is ``special" because of the peculiar critical point shift (\ref{wet})
which is somewhat anomalous and has attracted a lot of interest in
recent years \cite{parry,parry2,albano,binder2,kerle}.
In our opinion this point of view needs to be reconsidered; the mechanism
of critical point shift becomes clear only for nonzero gravity where
we find a scaling relation (\ref{newscal}) completely analogous to that 
expected for the capillary critical point (\ref{capill}).

Although they are currently restricted to two dimensions, DMRG techniques
provide accurate results for studying equilibrium properties of large 
systems. The fact that the DMRG accuracy is the best for open boundary 
conditions \cite{whitePRL,whitePRB} and that transfer matrices describe 
strips that are infinite along the transfer direction make the method 
best suited to study systems in contact with walls or with free surfaces.

We are grateful to J.O. Indekeu for suggesting us the topic of this 
Letter and for several discussions. 
Stimulating discussions with C. Boulter, R. Dekeyser, A.O. Parry and 
J. Rogiers are gratefully acknowledged. E.C. was supported by E.C. 
Human Capital and Mobility Programme N. CHBGCT940734. During his stay 
in Leuven A.D. was supported by the K.U. Leuven Research Fund (F/95/21).

\end{multicols}


\begin{references}
\bibitem{nota1} For instance, for an Ising model numerical computations
are typically restricted to $L \approx 15-20$.
\bibitem{nota2} We considered renormalized matrices with dimensions up to
$4096 \times 4096$, which corresponds to $32$ states kept in the DMRG 
language \cite{whitePRL}.
\bibitem{nota3} We restrict ourselves to $h_1 < J$ for which wetting
occurs at nonzero temperatures ($T_w \neq 0$).
\end{references}
\end{document}